\documentclass[pdflatex,sn-mathphys-num]{sn-jnl}

\usepackage{graphicx}
\usepackage{multirow}
\usepackage{booktabs}
\usepackage{amsmath}
\usepackage{algorithm}
\usepackage{algorithmic}
\usepackage{subcaption}

\begin{document}

\title[FL for Scalable AI in Heterogeneous HPC and Cloud]{Federated Learning Framework for Scalable AI in Heterogeneous HPC and Cloud Environments}

\author*[1]{\fnm{Sangam} \sur{Ghimire}}\email{1sangamghimire1@gmail.com}
\author[1]{\fnm{Paribartan} \sur{Timalsina}}\email{timalsinapari015@gmail.com}
\author[1]{\fnm{Nirjal} \sur{Bhurtel}}\email{bhurtelnirjal@gmail.com}
\author[1]{\fnm{Bishal} \sur{Neupane}}\email{neupanebishal2001@gmail.com}
\author[1]{\fnm{Bigyan} \sur{Byanju Shrestha}}\email{byanjubigyan9841@gmail.com}
\author[1]{\fnm{Subarna} \sur{Bhattarai}}\email{sbrnbhttr@gmail.com}
\author[1]{\fnm{Prajwal} \sur{Gaire}}\email{prajwalgaire617@gmail.com}
\author[1]{\fnm{Jessica} \sur{Thapa}}\email{thapajessi123456789@gmail.com}
\author[1]{\fnm{Sudan} \sur{Jha}}\email{jhasudan@ieee.org}

\affil[1]{\orgdiv{Department of Computer Science and Engineering}, \orgname{School of Engineering, Kathmandu University}, \orgaddress{\city{Dhulikhel}, \country{Nepal}}}

\abstract{
As the demand grows for scalable and privacy-aware AI systems, Federated Learning (FL) has emerged as a promising solution, allowing decentralized model training without moving raw data. At the same time, the combination of high-performance computing (HPC) and cloud infrastructure offers vast computing power but introduces new complexities, especially when dealing with heterogeneous hardware, communication limits, and non-uniform data. In this work, we present a federated learning framework built to run efficiently across mixed HPC and cloud environments. Our system addresses key challenges such as system heterogeneity, communication overhead, and resource scheduling, while maintaining model accuracy and data privacy. Through experiments on a hybrid testbed, we demonstrate strong performance in terms of scalability, fault tolerance, and convergence, even under non-Independent and Identically Distributed (non-IID) data distributions and varied hardware. These results highlight the potential of federated learning as a practical approach to building scalable Artificial Intelligence (AI) systems in modern, distributed computing settings.
}


\maketitle

\section{Introduction}

As AI models continue to grow in complexity and size, so does the demand for vast computational resources and access to large-scale distributed datasets. At the same time, growing concerns about data privacy, ownership, and regulatory compliance make it increasingly difficult to centralize data for training. FL has emerged as a promising paradigm for addressing these challenges, enabling the training of collaborative models across multiple data silos without requiring the raw data to leave its source. While FL has gained traction in mobile and edge environments, such as smartphones and IoT devices, its application in large-scale computing platforms like HPC clusters and cloud infrastructure remains underexplored.

Meanwhile, the convergence of HPC and cloud computing is reshaping the landscape of modern data-intensive applications. These hybrid environments combine the raw power and efficiency of HPC with the scalability and flexibility of the cloud, making them well-suited for training large AI models. However, this integration brings new challenges: heterogeneous hardware (e.g., Central Processing Units (CPUs), Graphics Processing Units (GPUs), Tensor Processing Units (TPUs)), inconsistent network performance, dynamic resource availability, and non-uniform data distributions across clients.

In this context, the deployment of federated learning across such mixed infrastructure is both a timely opportunity and a technical challenge. This paper explores how FL can be adapted and optimized to run efficiently across heterogeneous HPC and cloud environments, with a focus on scalability, system resilience, and performance under non-IID data conditions.

\section{Literature Review}

FL has rapidly grown as a research area since it was introduced by McMahan et al.~\cite{mcmahan2023communicationefficientlearningdeepnetworks}, with the goal of enabling decentralized model training while preserving user privacy. FL removes the need to centralize data by having clients train models locally and only share model updates. This is especially beneficial in settings where privacy, data security, or legal constraints prevent data sharing, such as healthcare, finance, and mobile systems.

Initial research in FL focused on edge computing environments where clients are resource-constrained devices such as smartphones or sensors~\cite{kairouz2021advancesopenproblemsfederated}. In these settings, major challenges include limited computation, high client churn, and extremely non-IID data distributions. Approaches such as Federated Averaging (FedAvg)~\cite{mcmahan2023communicationefficientlearningdeepnetworks}, FedProx~\cite{li2020federatedoptimizationheterogeneousnetworks}, and Federated Matched Averaging (FedMA)~\cite{wang2020federatedlearningmatchedaveraging} sought to improve convergence and personalization under such constraints. Systems like TensorFlow Federated~\cite{TensorFlowfederated}, FedML~\cite{he2020fedmlresearchlibrarybenchmark} and Flautim~\cite{barros2024flautim} were developed to provide flexible simulation and experimentation tools tailored for these edge use cases.

However, these systems generally assume relatively homogeneous client devices and do not scale effectively to enterprise-grade computing environments, such as data centers, cloud platforms, or HPC clusters. They are also not optimized for high-throughput interconnects, batch schedulers (like SLURM), or mixed-architecture setups (e.g., CPU, GPU, TPU nodes).

Several efforts have focused on benchmarking and simulating FL workloads. LEAF~\cite{caldas2019leafbenchmarkfederatedsettings} introduced a suite of datasets and evaluation metrics for real-world federated scenarios, while FedScale~\cite{lai2022fedscalebenchmarkingmodelperformance} expanded on this by providing a large-scale FL benchmarking platform that supports real-world training pipelines and system heterogeneity. Although these platforms help evaluate FL algorithm behavior at scale, they are primarily used for simulation, rather than actual deployment across hybrid infrastructures.

The idea of bringing FL to HPC and cloud environments has only recently begun to gain traction. Haus et al.~\cite{hpefederated} explored integrating FL with traditional HPC architectures, emphasizing challenges such as batch job scheduling and MPI-based communication. Similarly, the Adaptive Federated Learning framework introduced in~\cite{8664630} incorporates adaptive workload partitioning for federated learning on remote cloud platforms. However, these systems often operate in siloed infrastructure—either HPC-only or cloud-only—and do not address hybrid orchestration across both.

In contrast, recent work on cross-silo FL~\cite{bonawitz2019federatedlearningscaledesign} focuses on collaborative learning across data centers or enterprises. These scenarios involve more powerful clients with higher reliability and are more aligned with the settings of HPC and cloud systems. Yet, most existing implementations assume static infrastructure and overlook the dynamic nature of multi-cloud and hybrid HPC environments.

A major obstacle in deploying FL at scale is dealing with heterogeneity—not only in data, but in compute capabilities, memory, and network bandwidth. Approaches like Federated learning on non-IID data with Reinforcement Learning and quantifying historical contribution (FedRQ)~\cite{2024SPIE13224E..1CL} and FedBalancer~\cite{10.1145/3498361.3538917} introduce adaptive client selection and participation strategies to address system imbalance. Similarly, techniques such as gradient compression~\cite{lin2020deepgradientcompressionreducing}, gradient sparsification~\cite{9355797}, and adaptive synchronization~\cite{haddadpour2020localsgdperiodicaveraging} reduce communication overhead in bandwidth-constrained scenarios. Yet, incorporating these methods into a unified system that can operate across a wide spectrum of infrastructure types—HPC nodes with Infiniband, cloud instances with variable latency, or edge clients with intermittent connectivity—remains a significant engineering challenge.

In summary, the current FL landscape includes substantial work in algorithmic design, simulation platforms, and edge-device scenarios. However, few systems target practical deployment in environments where high-performance computing meets cloud infrastructure, which are increasingly common in research institutions, enterprises, and national labs. These hybrid setups introduce complex scheduling, heterogeneity, and data locality constraints that are not fully addressed by current FL frameworks. This paper aims to fill that gap by presenting a federated learning system built specifically for heterogeneous HPC–cloud environments, with contributions at both the system and algorithmic level: dynamic client orchestration, resource-aware scheduling, and robust model aggregation under non-IID and non-uniform resource conditions.

\section{Proposed System Architecture}

To support scalable and privacy-preserving federated learning in heterogeneous computing environments, we propose a modular and adaptable system architecture that integrates cloud-native and HPC resources under a unified framework. The design addresses core challenges such as client heterogeneity, resource variability, fault tolerance, and efficient communication—all while preserving data locality and minimizing orchestration overhead.

\begin{figure}[htbp]
    \centering
    \includegraphics[width=0.9\linewidth]{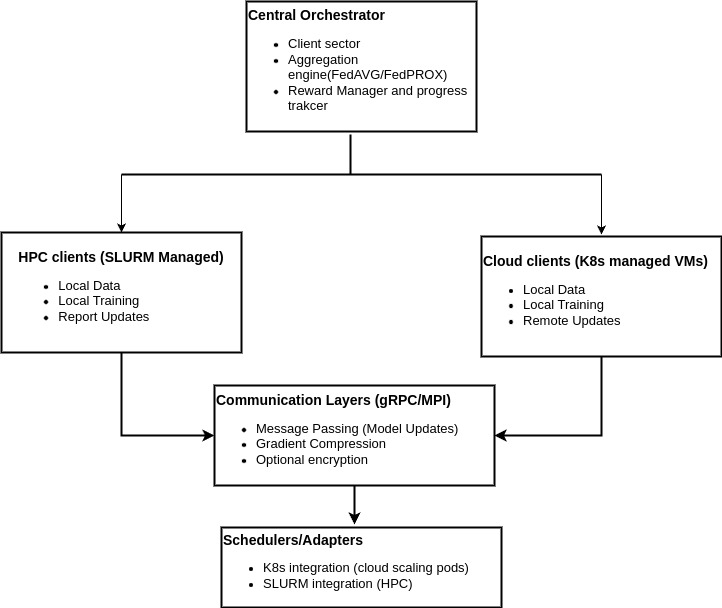}
    \caption{Architecture of the proposed federated learning framework.}
    \label{fig:architecture}
\end{figure}

The architecture consists of four primary components: \textbf{Central Orchestrator}, \textbf{Federated Clients (Workers)}, an optimized \textbf{Communication Layer}, and a dynamic \textbf{Scheduler Adapter}. Each component is designed to function independently yet integrate seamlessly, enabling operation across cloud instances, HPC clusters, or hybrid combinations of both. The proposed system architecture is shown in Fig.~\ref{fig:architecture}.

\subsection{Design Objectives}

Our system architecture is guided by the following key objectives:

\begin{itemize}
    \item \textbf{Scalability:} Efficiently support large-scale federated training across hundreds or thousands of distributed nodes with minimal coordination overhead.
    
    \item \textbf{Heterogeneity Awareness:} Adapt to variations in hardware configurations, network conditions, and runtime performance across diverse client nodes.
    
    \item \textbf{Data Privacy:} Ensure that raw data remains local to client nodes, in line with privacy-preserving principles and regulatory compliance.
    
    \item \textbf{Fault Tolerance:} Maintain robust training performance despite client dropouts, preemptions (e.g., in spot instances), or network disruptions.
\end{itemize}

\subsection{System Components}

\begin{itemize}
    \item \textbf{Central Orchestrator:}
    The central orchestrator is the core coordination unit responsible for managing the federated learning workflow. It is lightweight and stateless, and can be deployed on a cloud controller, an HPC login node, or a dedicated service. It carries out tasks like:
    dynamically selecting a subset of available clients for each round based on compute capacity, bandwidth, and past reliability; aggregating local model updates using strategies such as FedAvg~\cite{mcmahan2023communicationefficientlearningdeepnetworks} or FedProx~\cite{li2020federatedoptimizationheterogeneousnetworks}; and tracking training progress, managing round deadlines, and handling missing or delayed updates via fault-tolerant coordination logic.
\end{itemize}

\begin{itemize}
    \item \textbf{Communication Layer:}
    A flexible and efficient communication layer enables reliable messaging and bandwidth-aware transfer of updates. It supports protocols like gRPC (Remote Procedure Call) for cloud-native messaging and Message Passing Interface (MPI) for low-latency interconnects in HPC systems. For bandwidth efficiency it employs gradient compression, sparsification, and federated dropout to reduce communication overhead. To ensure secure data transmission and model training, the system can be extended with Transport Layer Security (TLS) encryption and secure aggregation.
\end{itemize}

\begin{itemize}
    \item \textbf{Scheduler Adapter:}
    The scheduler adapter provides an abstraction layer between the federated learning system and underlying resource managers. It integrates with SLURM in traditional HPC environments and handles job submission, monitoring, and resource allocation. For efficient management of containerized workloads, it supports dynamic pod orchestration and autoscaling through Kubernetes in cloud deployments. In addition, it supports hybrid coordination capabilities, facilitating scheduling across both HPC and cloud resources and enabling elastic, mixed-infrastructure setups.
\end{itemize}

\subsection{Workflow Overview}

Each training round begins with the orchestrator selecting clients based on resource profiling and availability. The global model is distributed, and clients train locally on their private data. After local training, clients send model updates back through the communication layer. The orchestrator aggregates these updates to generate a new global model. This process repeats iteratively until the model converges. The algorithm of the proposed workflow is given in Algorithm~\ref{algorithm}.

The modularity of this architecture enables flexible deployment, adaptive scaling, and robust performance even in the presence of heterogeneous hardware, non-uniform data, and unpredictable resource dynamics.

\begin{algorithm}
\caption{Federated Learning Training Procedure}
\label{algorithm}
\begin{algorithmic}[1]
    \REQUIRE Initial global model $M_0$, total rounds $R$, convergence threshold $\varepsilon$
    \ENSURE Final global model $M$

    \STATE Initialize round counter $r \leftarrow 0$
    \WHILE{$r < R$}
        \STATE Announce start of Round $r$
        \STATE Central orchestrator selects a subset of clients $\mathcal{C}_r$
        \STATE Send current global model $M_r$ to all clients in $\mathcal{C}_r$

        \FORALL{client $c \in \mathcal{C}_r$ \textbf{in parallel}}
            \STATE Client $c$ trains $M_r$ on local private data
            \STATE Client computes local model update $\Delta M_c$
            \STATE Client sends $\Delta M_c$ to the central orchestrator
        \ENDFOR

        \STATE Server aggregates updates: $\Delta M \leftarrow \text{Aggregate}(\{\Delta M_c \mid c \in \mathcal{C}_r\})$
        \STATE Update global model: $M_{r+1} \leftarrow M_r + \Delta M$

        \IF{\text{Converged}($M_r$, $M_{r+1}$, $\varepsilon$)}
            \STATE \textbf{return} $M_{r+1}$
        \ENDIF

        \STATE $r \leftarrow r + 1$
    \ENDWHILE

    \RETURN $M_r$
\end{algorithmic}
\end{algorithm}

\section{Heterogeneity-Aware Optimization}

Heterogeneity in FL arises not only from data (non-IID distributions) but also from system-level differences such as compute power, memory, network bandwidth, and availability across participating nodes. These disparities are especially pronounced in mixed HPC and cloud environments, where client capabilities can vary widely. In this section, we describe the optimization techniques incorporated into our system to address these challenges and ensure robust and efficient training.

\subsection{Adaptive Client Selection}

Selecting the right subset of clients for each training round is critical for maintaining both fairness and efficiency in heterogeneous settings. Our system implements an \textbf{adaptive client selection mechanism} that takes into account several factors such as resource profiling, performance history, and load balancing. 

Through resource profiling, clients are periodically benchmarked for available CPU/GPU capacity, memory, and network latency. For performance history, the orchestrator maintains a record of successful participation, update quality, and completion time for each client. Under load balancing, underperforming or slower nodes may be temporarily excluded to avoid slowing down the round or causing stale updates. By dynamically adjusting the client pool per round, the system avoids bottlenecks and maintains training momentum, even when participating nodes vary significantly in capability.

\subsection{Straggler Mitigation}

In federated learning, \textit{stragglers}—clients that take disproportionately long to complete local training—can delay or degrade the effectiveness of global aggregation. To handle this, we employ two complementary techniques:

\begin{itemize}
    \item \textbf{Deadline-Based Cutoff:} Each round has a configurable time budget. Clients that fail to report updates within the deadline are skipped for that round, reducing the impact of delays~\cite{goren2025adaptivedeadlinebatchlayered}.
    
    \item \textbf{Partial Aggregation:} Rather than waiting for all selected clients, the orchestrator aggregates updates from the fastest $k$ clients (where $k$ is tunable), ensuring timely global updates without compromising overall convergence~\cite{LIU2021108468}.
\end{itemize}

These approaches help mitigate the impact of slow or overloaded nodes, particularly useful in environments with spot instances or shared HPC queues.

\subsection{Communication-Efficient Updates}

Communication is often a limiting factor in federated learning, especially across geographically dispersed cloud regions or bandwidth-constrained HPC interconnects. Our framework incorporates several techniques to reduce communication overhead. One of them is gradient quantization, which reduces the precision of model updates before transmission, minimizing bandwidth without significantly affecting accuracy. Similarly, using update sparsification, clients transmit only a subset of the most important gradients (e.g., top-$k$ by magnitude), further reducing data size. Furthermore, we employ federated dropout where clients train and communicate only partial models by randomly dropping out neurons or layers, decreasing both compute and communication loads. These techniques are especially valuable when federated clients operate over variable or costly network links, such as cloud egress bandwidth.

\subsection{Robust Aggregation under Non-IID Data}

Our system supports multiple aggregation methods tailored for non-IID client data distributions. In addition to the standard FedAvg~\cite{mcmahan2023communicationefficientlearningdeepnetworks}, we implement FedProx, which adds a proximal term to local objectives to prevent client models from drifting too far from the global model, improving stability in heterogeneous data conditions~\cite{li2020federatedoptimizationheterogeneousnetworks}, and weighted aggregation that assigns dynamic weights to each client update based on local data size, training loss, or gradient variance, improving fairness and convergence. 

This combination of algorithmic robustness and system-level optimizations ensures that training remains effective even under significant data and system heterogeneity.

\begin{table*}[htbp]
\centering
\small
\caption{Summary of Heterogeneity-Aware Optimization Techniques}
\label{tab:optimizations}
\begin{tabular}{p{3.5cm} p{3.2cm} p{8.5cm}}
\toprule
\textbf{Challenge} & \textbf{Technique} & \textbf{Description} \\
\midrule
System Heterogeneity & Adaptive Client Selection & Clients selected based on compute capability, availability, and historical performance. \\
Straggler Clients & Deadline-Based Cutoff & Late clients are excluded from aggregation if they miss the round deadline. \\
& Partial Aggregation & Aggregation proceeds after receiving updates from the fastest $k$ clients. \\
Communication Overhead & Gradient Quantization & Model updates are compressed by reducing numerical precision (e.g., 16-bit, 8-bit). \\
& Update Sparsification & Only top-$k$ most significant gradients are sent, reducing payload size. \\
& Federated Dropout & Clients train and transmit reduced model subsets to lower communication cost. \\
Data Heterogeneity (Non-IID) & FedProx & Adds a regularization term to stabilize training with divergent local updates. \\
& Weighted Aggregation & Updates weighted by factors like local dataset size or training loss. \\
\bottomrule
\end{tabular}
\end{table*}

\section{Experimental Evaluation}

To validate the effectiveness, scalability, and robustness of our proposed federated learning framework, we conducted a series of experiments in a hybrid computing environment comprising both HPC and cloud resources. This section describes our experimental setup, datasets, evaluation metrics, and key findings.

\subsection{Experimental Setup}

Our testbed consists of a hybrid cluster with 30 virtual machines (VMs) on Amazon Web Services Elastic Compute Cloud (AWS EC2), including both GPU-enabled (p3.2xlarge) and CPU-only (t3.large) instances as cloud nodes, and 30 compute nodes managed by SLURM, equipped with NVIDIA Quadro RTX~6000 GPUs with 24\,GB VRAM and 4608 CUDA cores, as well as CPU-only nodes as HPC nodes.

For the configuration of the framework we use a communication backend that utilizes gRPC for cloud environments and MPI for HPC setups. Algorithms such as FedAvg and FedProx are supported and 20 clients are randomly selected in each training round, while there are 100 training rounds for the entire training process. Each client performs 5 local epochs of training before sending its updates for aggregation.

\subsection{Datasets}

We evaluated our framework on the following benchmark datasets, each partitioned to simulate non-IID data distributions across clients:
\begin{itemize}
    \item \textbf{CIFAR-10:} Image classification with 10 classes. Each client receives samples from only 2–3 classes.
    \item \textbf{Shakespeare (LEAF):} A character-level language modeling task on dialogue data.
    \item \textbf{MedMNIST:} Medical image classification task simulating privacy-sensitive healthcare settings.
\end{itemize}

\subsection{Evaluation Metrics}

We used the following metrics to evaluate system performance:
\begin{itemize}
    \item \textbf{Model Accuracy:} Test accuracy on a centralized evaluation dataset.
    \item \textbf{Convergence Speed:} Number of rounds to reach a defined accuracy threshold.
    \item \textbf{Training Time:} End-to-end wall-clock time per round and total training time.
    \item \textbf{Communication Overhead:} Average data transmitted per client per round.
    \item \textbf{Fault Tolerance:} Model accuracy under simulated client dropouts or preemptions.
\end{itemize}

\subsection{Results Summary}

\begin{figure}[htbp]
    \centering
    \begin{subfigure}{0.32\textwidth}
        \centering
        \includegraphics[width=\linewidth]{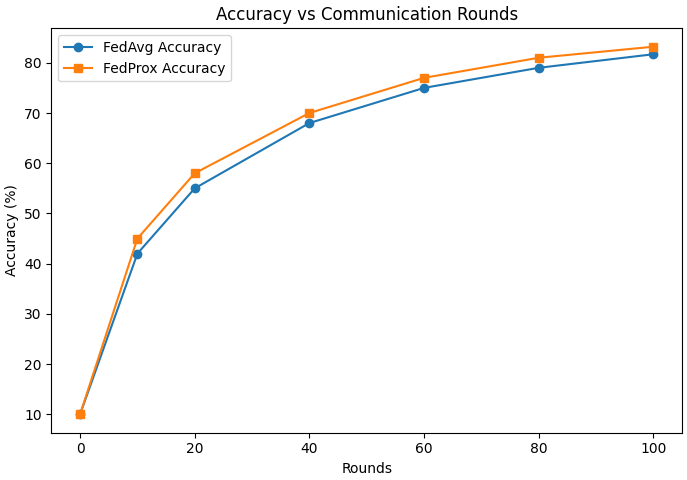}
        \caption{CIFAR-10}
        \label{fig:accuracy_cifar}
    \end{subfigure}
    \hfill
    \begin{subfigure}{0.32\textwidth}
        \centering
        \includegraphics[width=\linewidth]{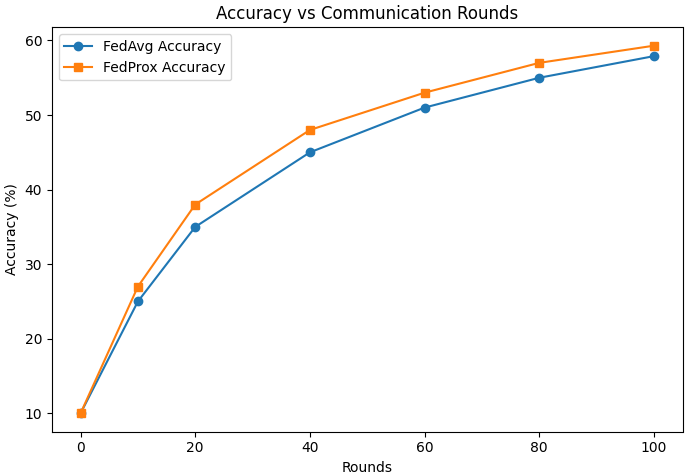}
        \caption{Shakespeare}
        \label{fig:accuracy_shakespeare}
    \end{subfigure}
    \hfill
    \begin{subfigure}{0.32\textwidth}
        \centering
        \includegraphics[width=\linewidth]{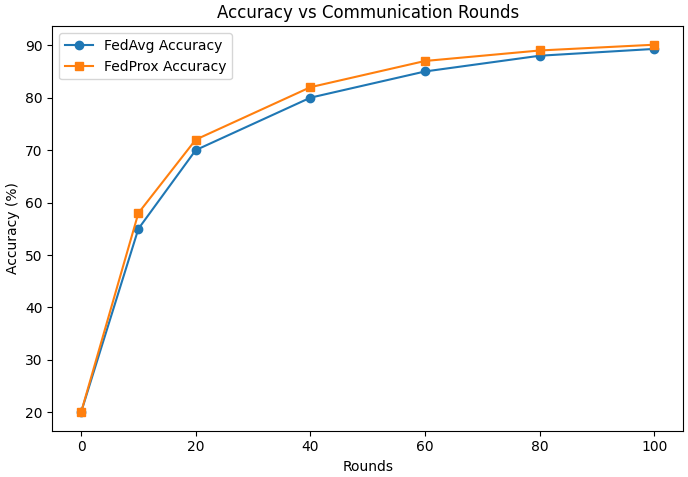}
        \caption{MedMNIST}
        \label{fig:accuracy_medmnist}
    \end{subfigure}
    \caption{Accuracy comparison of FedAvg and FedProx across different datasets under non-IID settings.}
    \label{fig:accuracy}
\end{figure}

\textbf{Accuracy and Convergence:} Our system achieved competitive model accuracy across all datasets. Under non-IID conditions, FedProx showed improved convergence stability over FedAvg. The dataset-specific result plots are shown in Fig.~\ref{fig:accuracy} and summarized numerically in Table~\ref{tab:accuracy}.

\begin{table}[htbp]
\centering
\caption{Accuracy Comparison Between FedAvg and FedProx}
\label{tab:accuracy}
\begin{tabular}{lcc}
\toprule
\textbf{Dataset} & \textbf{FedAvg} & \textbf{FedProx} \\
\midrule
CIFAR-10   & 81.7\% & 83.2\% \\
Shakespeare & 57.9\% & 59.3\% \\
MedMNIST   & 89.3\% & 90.1\% \\
\bottomrule
\end{tabular}
\end{table}

\textbf{Scalability:} Increasing the number of clients from 10 to 60 showed near-linear improvements in training throughput, achieving a 4.6$\times$ speedup in total training time, as tabulated in Table~\ref{tab:scalability}.

\begin{table}[htbp]
\centering
\caption{Scalability Analysis with Varying Number of Clients}
\label{tab:scalability}
\begin{tabular}{ccc}
\toprule
\textbf{Number of Clients} & \textbf{Total Time (min)} & \textbf{Speedup ($\times$)} \\
\midrule
10  & 100 & 1.00 \\
20  & 58  & 1.72 \\
30  & 43  & 2.32 \\
40  & 33  & 3.03 \\
50  & 27  & 3.70 \\
60  & 22  & 4.55 \\
\bottomrule
\end{tabular}
\end{table}

\textbf{Straggler Resilience:} With 20\% simulated client dropouts per round, the final accuracy dropped by less than 1.8\%, demonstrating strong fault tolerance.

\textbf{Communication Efficiency:} Gradient quantization and sparsification reduced the average communication volume by about 65\% without significant accuracy degradation, as shown in Table~\ref{tab:compression}.

\begin{table}[htbp]
\centering
\caption{Communication Volume Per Round With and Without Compression}
\label{tab:compression}
\begin{tabular}{ccc}
\toprule
\textbf{Round} & \textbf{No Compression (MB)} & \textbf{With Compression (MB)} \\
\midrule
1  & 45 & 16 \\
2  & 44 & 15 \\
3  & 43 & 14 \\
4  & 44 & 15 \\
5  & 43 & 14 \\
6  & 42 & 14 \\
7  & 44 & 15 \\
8  & 43 & 14 \\
9  & 42 & 13 \\
10 & 43 & 14 \\
\bottomrule
\end{tabular}
\end{table}

\subsection{Ablation Studies}

To assess the contribution of each optimization, we conducted ablation experiments disabling one component at a time. Without adaptive client selection, we observed a 12\% increase in average round duration. Without communication compression there was a 70\% increase in bandwidth usage. Similarly, without straggler mitigation we saw 15–20\% longer training time to reach 80\% accuracy. These results confirm that heterogeneity-aware optimizations are essential to maintain both efficiency and reliability in real-world deployments.

\section{Discussion, Limitations and Future Work}

Our experimental results demonstrate that federated learning can be made both scalable and practical in heterogeneous environments by carefully addressing system-level challenges. The integration of adaptive scheduling, communication-efficient updates, and fault tolerance mechanisms enables reliable training across cloud and HPC infrastructure—even under non-IID data conditions.

The ability to run federated learning over hybrid compute environments has strong implications for industries where data is siloed across institutions or departments (e.g., healthcare, finance, scientific research). Our system design shows that FL is no longer limited to edge devices or simulated environments but can be deployed at enterprise and cluster scale.

Heterogeneity is not a flaw; it is a feature. Rather than treating hardware diversity as a limitation, our system embraces it. By tailoring workloads based on node capability and using adaptive aggregation, slower or lower-power nodes can still contribute without holding back global progress.

While compute is abundant, bandwidth often is not—especially in cloud-based federated setups. Our results show that communication optimizations such as quantization and sparsification significantly reduce training costs without affecting model quality, making FL more feasible in real-world deployments.

Training in distributed environments means expecting failure. Our system showed strong tolerance to client dropouts, but future work must address adversarial behavior and secure aggregation to make FL trustworthy at scale.

Despite promising results, some challenges remain. The current system assumes relatively static client availability during each round. Highly dynamic workloads (e.g., serverless or elastic computing) would benefit from runtime-aware orchestration. Additionally, our experiments focus on moderate-scale clusters; scaling to hundreds or thousands of nodes introduces new coordination and aggregation bottlenecks that require further study.

Furthermore, several directions remain open for exploration:

\begin{itemize}
    \item \textbf{Secure aggregation:} Incorporating cryptographic techniques (e.g., differential privacy, homomorphic encryption) to enhance data confidentiality.
    
    \item \textbf{Elastic orchestration:} Designing a more dynamic scheduling engine that reacts to real-time availability and pricing of cloud and HPC nodes.
    
    \item \textbf{Federated inference:} Extending the current training framework to support decentralized, real-time inference across the client network.
    
    \item \textbf{Integration with foundation models:} Adapting the system to fine-tune and serve large-scale pre-trained models (e.g., Large Language Models (LLMs)) in federated scenarios.
\end{itemize}

\section{Conclusion}

This paper presented a federated learning framework designed to operate in heterogeneous high-performance computing and cloud environments. By combining adaptive client selection, communication-efficient updates, and fault-tolerant aggregation strategies, our system achieves robust performance even under challenging conditions such as system variability and non-IID data.

Through empirical evaluations on hybrid cloud–HPC testbeds, we showed that the proposed approach delivers strong accuracy, scalability, and resilience. The results underline that federated learning is not only viable but advantageous in multi-infrastructure environments where data privacy and system diversity are critical.

Our work represents a step toward scalable, private, and intelligent distributed learning systems that can thrive in the heterogeneous computing ecosystems of the future.

\bibliography{sn-bibliography}

\end{document}